\documentclass[twocolumn,showpacs,prl,preprintnumbers,superscriptaddress]{revtex4}
\pdfoutput=1

\usepackage{graphicx}
\usepackage{epsfig}
\usepackage{xspace}
\usepackage{dcolumn}
\usepackage{multirow}
\usepackage{longtable}

\newcolumntype{d}[1]{D{.}{\cdot}{#1}}
\newcolumntype{.}{D{.}{.}{-1}}
\newcolumntype{,}{D{,}{,}{2}}
\begin{document}

\title{ Search for Differences in Oscillation Parameters for Atmospheric Neutrinos and Antineutrinos at Super-Kamiokande }

\newcommand{\AFFicrr}{\affiliation{Kamioka Observatory, Institute for Cosmic Ray Research, University of Tokyo, Kamioka, Gifu 506-1205, Japan}}
\newcommand{\AFFkashiwa}{\affiliation{Research Center for Cosmic Neutrinos, Institute for Cosmic Ray Research, University of Tokyo, Kashiwa, Chiba 277-8582, Japan}}
\newcommand{\AFFipmu}{\affiliation{Institute for the Physics and
Mathematics of the Universe, University of Tokyo, Kashiwa, Chiba
277-8582, Japan}}
\newcommand{\AFFuam}{\affiliation{Department of Theoretical Physics, University Autonoma Madrid, 28049 Madrid, Spain }}
\newcommand{\AFFbu}{\affiliation{Department of Physics, Boston University, Boston, MA 02215, USA}}
\newcommand{\AFFbnl}{\affiliation{Physics Department, Brookhaven National Laboratory, Upton, NY 11973, USA}}
\newcommand{\AFFucd}{\affiliation{Department of Physics, University of California, Davis, Davis, CA 95616, USA}}
\newcommand{\AFFuci}{\affiliation{Department of Physics and Astronomy, University of California, Irvine, Irvine, CA 92697-4575, USA }}
\newcommand{\AFFcsu}{\affiliation{Department of Physics, California State University, Dominguez Hills, Carson, CA 90747, USA}}
\newcommand{\AFFcnm}{\affiliation{Department of Physics, Chonnam National University, Kwangju 500-757, Korea}}
\newcommand{\AFFduke}{\affiliation{Department of Physics, Duke University, Durham NC 27708, USA}}
\newcommand{\AFFgmu}{\affiliation{Department of Physics, George Mason University, Fairfax, VA 22030, USA }}
\newcommand{\AFFgifu}{\affiliation{Department of Physics, Gifu University, Gifu, Gifu 501-1193, Japan}}
\newcommand{\AFFuh}{\affiliation{Department of Physics and Astronomy, University of Hawaii, Honolulu, HI 96822, USA}}
\newcommand{\AFFkanagawa}{\affiliation{Physics Division, Department of Engineering, Kanagawa University, Kanagawa, Yokohama 221-8686, Japan}}
\newcommand{\AFFkek}{\affiliation{High Energy Accelerator Research Organization (KEK), Tsukuba, Ibaraki 305-0801, Japan }}
\newcommand{\AFFkobe}{\affiliation{Department of Physics, Kobe University, Kobe, Hyogo 657-8501, Japan}}
\newcommand{\Pkobe}{Department of Physics, Kobe University, Kobe, Hyogo 657-8501, Japan}
\newcommand{\AFFkyoto}{\affiliation{Department of Physics, Kyoto University, Kyoto, Kyoto 606-8502, Japan}}
\newcommand{\AFFumd}{\affiliation{Department of Physics, University of Maryland, College Park, MD 20742, USA }}
\newcommand{\AFFmit}{\affiliation{Department of Physics, Massachusetts Institute of Technology, Cambridge, MA 02139, USA}}
\newcommand{\AFFmiyagi}{\affiliation{Department of Physics, Miyagi University of Education, Sendai, Miyagi 980-0845, Japan}}
\newcommand{\AFFnagoya}{\affiliation{Solar Terrestrial Environment
Laboratory, Nagoya University, Nagoya, Aichi 464-8602, Japan}}
\newcommand{\AFFsuny}{\affiliation{Department of Physics and Astronomy, State University of New York, Stony Brook, NY 11794-3800, USA}}
\newcommand{\AFFniigata}{\affiliation{Department of Physics, Niigata University, Niigata, Niigata 950-2181, Japan }}
\newcommand{\AFFokayama}{\affiliation{Department of Physics, Okayama University, Okayama, Okayama 700-8530, Japan }}
\newcommand{\AFFosaka}{\affiliation{Department of Physics, Osaka University, Toyonaka, Osaka 560-0043, Japan}}
\newcommand{\AFFseoul}{\affiliation{Department of Physics, Seoul National University, Seoul 151-742, Korea}}
\newcommand{\AFFshizuokasc}{\affiliation{Department of Informatics in
Social Welfare, Shizuoka University of Welfare, Yaizu, Shizuoka, 425-8611, Japan}}
\newcommand{\AFFshizuoka}{\affiliation{Department of Systems Engineering, Shizuoka University, Hamamatsu, Shizuoka 432-8561, Japan}}
\newcommand{\AFFskk}{\affiliation{Department of Physics, Sungkyunkwan University, Suwon 440-746, Korea}}
\newcommand{\AFFtohoku}{\affiliation{Research Center for Neutrino Science, Tohoku University, Sendai, Miyagi 980-8578, Japan}}
\newcommand{\AFFtokyo}{\affiliation{The University of Tokyo, Bunkyo, Tokyo 113-0033, Japan }}
\newcommand{\AFFtokai}{\affiliation{Department of Physics, Tokai University, Hiratsuka, Kanagawa 259-1292, Japan}}
\newcommand{\AFFtit}{\affiliation{Department of Physics, Tokyo Institute
for Technology, Meguro, Tokyo 152-8551, Japan }}
\newcommand{\AFFtsinghua}{\affiliation{Department of Engineering Physics, Tsinghua University, Beijing, 100084, China}}
\newcommand{\AFFwarsaw}{\affiliation{Institute of Experimental Physics, Warsaw University, 00-681 Warsaw, Poland }}
\newcommand{\AFFuw}{\affiliation{Department of Physics, University of Washington, Seattle, WA 98195-1560, USA}}
\newcommand{\AFFkmiop}{\affiliation{Kobayashi-Maskawa Institute for the Origin of Particle
and the Universe,  Nagoya University, Nagoya, Aichi 464-8602, Japan}}

\AFFicrr
\AFFkashiwa
\AFFipmu
\AFFbu
\AFFbnl
\AFFuci
\AFFcsu
\AFFcnm
\AFFduke
\AFFgifu
\AFFuh
\AFFkanagawa
\AFFkek
\AFFkobe
\AFFkyoto
\AFFmiyagi
\AFFnagoya
\AFFsuny
\AFFniigata
\AFFokayama
\AFFosaka
\AFFseoul
\AFFshizuoka
\AFFshizuokasc
\AFFskk
\AFFtokai
\AFFtokyo
\AFFtsinghua
\AFFwarsaw
\AFFuw
%

\author{K.~Abe}
\author{Y.~Hayato}
\AFFicrr
\AFFipmu
\author{T.~Iida}
\author{M.~Ikeda}
\author{K.~Iyogi} 
\AFFicrr
\author{J.~Kameda}
\author{Y.~Koshio}
\AFFicrr
\AFFipmu
\author{Y.~Kozuma} 
\AFFicrr
\author{M.~Miura} 
\author{S.~Moriyama} 
\author{M.~Nakahata} 
\author{S.~Nakayama} 
\author{Y.~Obayashi} 
\author{H.~Sekiya} 
\author{M.~Shiozawa} 
\author{Y.~Suzuki} 
\author{A.~Takeda} 
\AFFicrr
\AFFipmu
% ...

\collaboration{The Super-Kamiokande Collaboration}
\noaffiliation

\date{\today}

\begin{abstract}

   We present a search for differences in the oscillations of antineutrinos and
   neutrinos in the Super-Kamiokande -I, -II, and -III atmospheric neutrino sample.
   Under a two-flavor disappearance model with separate mixing parameters between 
   neutrinos and antineutrinos, we find no evidence for a difference in oscillation parameters. 
   Best fit antineutrino mixing is found to be at 
   ($\Delta \bar m^{2}$, $\mbox{sin}^{2} 2\bar{\theta}) = ( 2.0\times 10^{-3} \mbox{eV}^{2} , 1.0)$ 
   and is consistent with the overall Super-K measurement.  

\end{abstract}


\maketitle


As the parameters outlining the standard neutrino oscillation framework 
become increasingly well known, searches for sub-leading and possibly 
symmetry-breaking effects become possible. If the value of $\theta_{13}$ is non-zero,
for instance, it becomes possible to search for CP-violation effects in the neutrino 
system via differences in the oscillation probabilities of neutrinos and antineutrinos. 
In this paper we consider the possibility that the survival probability 
$\mbox{P}(\nu_{\mu} \rightarrow \nu_{\mu} )$ is governed by a different 
mass splitting or mixing angle compared to $\mbox{P}(\bar \nu_{\mu} \rightarrow \bar \nu_{\mu} ).$
This is not considered in most oscillation studies as the mass splitting and 
mixing angle are expected to be identical for neutrinos and antineutrinos by CPT symmetry. 
An inequality of these probabilities, in the absence of matter effects, could signal new physics.
For atmospheric muon neutrino disappearance, which is predominantly oscillation into tau neutrinos, the
matter effect is expected to be small and the mixing parameters for muon neutrino and antineutrino 
disappearance appearance should be the same.
The MINOS experiment, which is sensitive to neutrino oscillations at the atmospheric 
scale, and which can determine the sign of muons by magnetic bending,
has observed antineutrino disappearance~\cite{Adamson:2011fa} at a best-fit value of $\Delta m^{2}$ nearly 50\%
larger than previous measurements made using neutrinos~\cite{Hosaka:2006zd,ashie:2005ik,K2K06:prd,Minos08}.
Though not in the realm of atmospheric mixing, the MiniBooNE experiment has similarly 
observed a discrepancy between its neutrino~\cite{AguilarArevalo:2007it} 
and antineutrino~\cite{AguilarArevalo:2010wv} data.
Therefore, further tests of differences between neutrinos and antineutrinos 
using atmospheric data are well motivated.

Super-Kamiokande (Super-K, SK), described below, 
cannot distinguish $\nu$ from $\bar \nu$ on 
an event by event basis so potential differences in their oscillations  
would appear in the atmospheric neutrino sample in a statistical way.
Notably, the neutrino and antineutrino cross sections differ 
by a factor of two to three depending on the neutrino energy.
The ratio of the $\nu$ and $\bar \nu$ atmospheric fluxes is similarly energy dependent~\cite{honda}.
For these reasons, even in the absence of CPT-violating oscillations 
the relative numbers of each species are expected to differ. 
Kinematic considerations can also induce differences in the products 
of neutrinos and antineutrino reactions, enhancing the sample purity of one or the other.
For instance, the absorption of $\pi^{-}$ on $^{16}\mbox{O}$ nuclei in water
tends to enrich the neutrino component of samples that are sub-divided  
based on their number of decay electrons.
If CPT-violating oscillations are present in the data, the distortion of the 
zenith angle distribution characteristic of $\nu_{\mu}$ disappearance 
would appear at different energies and path lengths 
(different oscillation frequencies) between neutrinos and antineutrinos. 
Since Super-K can only observe the total distribution, a potential 
signal would appear as a distortion consistent with the 
composition of separately oscillated spectra.
In this context, antineutrinos, which have the smaller cross section, 
are expected to provide a weaker oscillation constraint than neutrinos.

In this Letter we consider \textit{ad-hoc} CPT-violating oscillations
by testing separate two-neutrino disappearance models for neutrinos and antineutrinos: 
\begin{eqnarray}
\mbox{P}(\nu_{\mu} \rightarrow \nu_{\mu} ) &=& 1-\mbox{sin}^{2} 2\theta 
                       \mbox{sin}\left( \frac{\Delta m^{2}L}{4E} \right) \nonumber \\
\mbox{P}(\bar \nu_{\mu} \rightarrow \bar \nu_{\mu} ) &=& 
              1-\mbox{sin}^{2} 2\bar \theta 
                        \mbox{sin}\left( \frac{\Delta \bar m^{2}L }{4E}\right),  
\label{eq:prob}
\end{eqnarray}
\noindent where $L$ is the neutrino path-length and $E$ is the neutrino energy. 
 In the presence of matter, additional neutrino-electron scattering induces 
a CPT-violating-like difference between the neutrino and antineutrino 
survival probabilities~\cite{Mikheev:1986gs,Wolfenstein:1977ue}, particularly when $\theta_{13}$ is non-zero.
Recent data suggest~\cite{Chooz03, Wendell:2010sd, Abe:2011sj} that $\theta_{13}$ 
is small and therefore matter-induced corrections to Eqn.~\ref{eq:prob} 
are expected to be sub-dominant. 
Changes to the fit results induced by a three-flavor 
treatment are briefly considered below. 
Although the presence of both matter and solar mixing terms is expected 
to drive $\nu_{\mu} \rightarrow \nu_{e}$ transitions below about 1~GeV 
even if $\theta_{13} = 0$, the oscillation frequency in this domain 
is high enough that the effects are averaged out by the detector resolution. 
Since the dominant atmospheric disappearance effect is seen 
at higher energies, a two-neutrino scheme is the focus of this study. 

\begin{figure*}[htb]
  \begin{minipage}{5.0in}
      \includegraphics[width=5.0in,keepaspectratio=true,type=pdf,ext=.pdf,read=.pdf]{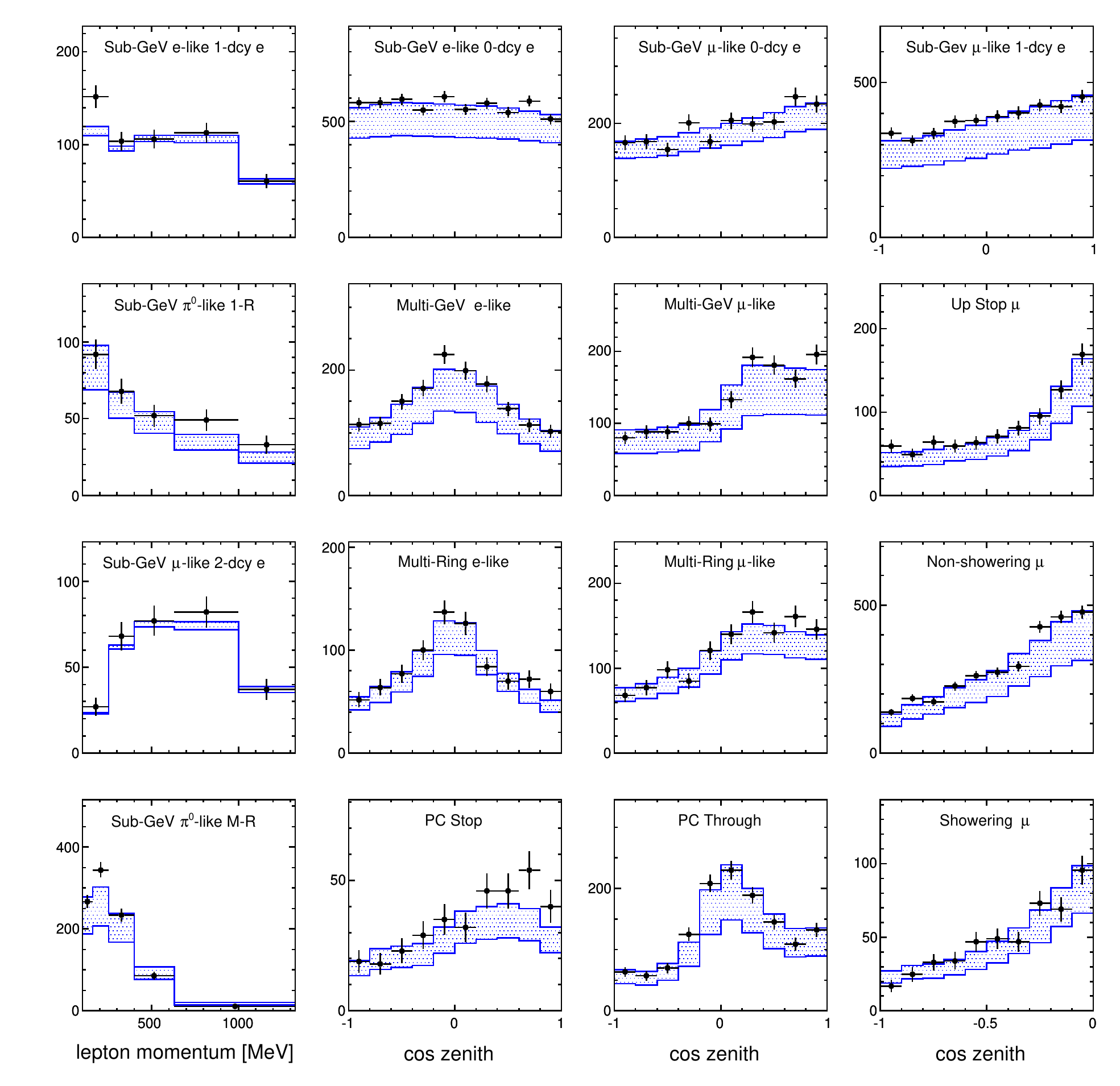}
  \end{minipage}
    \caption{ SK-I+II+III lepton momentum (first column) and zenith angle 
              distributions of the event samples used in the analysis. 
              Black dots represent the data with statistical errors,
              the histogram is the oscillated MC expectation at the best fit point 
              with the shaded region showing the antineutrino composition.  
            }
  \label{fig:cpt.zenith}
\end{figure*}

Super-Kamiokande is a water Cherenkov detector located in Japan's Gifu prefecture 
and situated at a depth of 2700 meters water equivalent.
It is comprised of two concentric, optically separated cylinders: an inner detector (ID) viewing 
a 22.5 kton fiducial volume and an outer detector (OD) used primarily as a veto. 
During the first run of the detector, SK-I, the walls of the ID were lined with 
11,146 inward-facing 20-inch photomultiplier tubes (PMTs). The two subsequent 
run periods, SK-II and SK-III, had 5,182 and 11,129 ID PMTs, respectively 
and the PMTs have been encased in fiber-reinforced plastic shells.  The OD has been 
instrumented with 1,885 outward-facing 8-inch PMTs throughout.
More details on the detector and its calibration may be found in~\cite{fukuda:2002uc}.

The atmospheric neutrino data are divided into three categories. 
Fully contained (FC) events deposit all of their light in the ID,
partially contained (PC) events additionally have an exiting particle that 
deposits energy in the OD, and upward-going muon (Up-$\mu$) events are produced
by neutrino interactions in the rock beneath the detector. Up-$\mu$ events
are required to have a minimum pathlength of seven meters 
and are classified as stopping or through-going.
Through-going Up-$\mu$ events are further sub-divided 
into ``showering'' and ``non-showering'' based on~\cite{Desai08}. 

The present analysis uses data accumulated during the first three SK run periods. 
SK-I (1996 to 2001) FC and PC events correspond to 1489 live-days with 1646 days 
of Up-$\mu$ livetime.
SK-II (2002 to 2005) had 799 FC/PC and 828 live-days of Up-$\mu$ events. 
The FC and PC livetimes during SK-III (2005-2007) were 518 days and that for 
Up-$\mu$ was 636 days. 

During the analysis, FC and PC events are further divided. 
Fully contained events are separated into sub-GeV ( $\mbox{E}_{vis} < 1.33$ GeV )
and multi-GeV ( $\mbox{E}_{vis} > 1.33$ GeV ). These samples are then separated 
based on their number of reconstructed Cherenkov rings into single- 
and multi-ring topologies. Pattern-identification of single-ring events is used to 
separate them into $e$-like and $\mu$-like categories. This technique is 
applied to the multi-ring sample using the most energetic Cherenkov ring.  
The sub-GeV single-ring e-like and $\mu$-like samples are also divided based 
upon their number of decay-electrons. A $\pi^{0}$-like sample is also 
extracted from the single-ring e-like events~\cite{Wendell:2010sd}. 
Partially contained events are separated into 
``OD stopping'' and ``OD through-going'' categories based on the amount of 
Cherenkov light observed in the OD at the exit point.

Since the physical configuration of the detector and its reconstruction 
performance varies among the SK run periods, separate 500 year MC samples are used for each. 
A ``pulled'' $\chi^{2}$~\cite{Lisi02} based on a 
Poisson probability distribution is used to compare the data against the MC:
%
\begin{widetext}
\begin{eqnarray}
\chi^{2}  &=& 2 \displaystyle \sum_{n} \left( 
              \displaystyle \sum_{i}E_{n}^{SKi}( 1 + \displaystyle \sum_{j} f^{j}_{n} \epsilon_{j} )
              - \displaystyle \sum_{i}\mathcal{O}_{n}^{SKi} 
              + \displaystyle \sum_{i} \mathcal{O}_{n}^{SKi}  
              \ln \frac{ \sum_{i}  \mathcal{O}_{n}^{SKi} }
                       { \sum_{i} E_{n}^{SKi}( 1 +  \sum_{j} f^{j}_{n} \epsilon_{j} ) } 
               \right)
             + \displaystyle \sum_{k} \left( \frac{ \epsilon_{k} }{ \sigma_{k} } \right)^{2}. 
\label{eq:fullchi}
\end{eqnarray}
\end{widetext}

\noindent 
In this equation $n$ indexes the data bins, $E_{n}^{SKi}$ is the MC expectation for SK-$i$, 
and $\mathcal{O}_{n}^{SKi}$ is the number of observed events in the $n^{th}$ bin during SK-$i$.
The effect of the $i^{th}$ systematic error is introduced via the error parameter $\epsilon_{i}$ and 
$f^{i}_{n}$, where the latter is the fractional change in the MC expectation of bin $n$ introduced by a 
1-sigma shift in its systematic error, $\sigma_{i}$. 
The data and MC are divided into 420 bins for each of the SK run periods when computing these
systematic errors, but are later merged as above to ensure the stability of the fit function against 
sparsely populated bins. In total, 420 bins are used to compute the value of $\chi^{2}.$  

Equation~(\ref{eq:fullchi}) is minimized with respect to the $\epsilon_{i}$ 
according to $\frac{ \partial \chi^{2}}{ \partial \epsilon_{i}} = 0$, yielding
a set of linear equations in $\epsilon_{i}$ that are solved iteratively~\cite{Lisi02}. 
Following this procedure a $\chi^{2}$ value is computed for each point in the oscillation parameter space.
The global minimum $\chi^{2}$ is defined as the analysis' best fit point. 

\begin{table}[htbp]
   \begin{center}
      \begin{tabular}{lccc}
      \hline
      \hline
      \\[0.1mm]
      Parameter                       & Best Fit                &  90\% C.L.         &  Three-Flavor  \\
      \hline
      \\[0.05mm]
 $\Delta m^{2} (\mbox{eV}^2)$         &  $2.1\times 10^{-3}$   &   [ 1.7 , 3.0 ]$\times 10^{-3}$    
                                                               &   [ 1.7 , 3.3 ]$\times 10^{-3}$ \\ 
 $\Delta \bar m^{2} (\mbox{eV}^2)$    &  $2.0\times 10^{-3}$   &   [ 1.3 , 4.0 ]$\times 10^{-3}$     
                                                               &   [ 1.2 , 4.0 ]$\times 10^{-3}$ \\ 
 $\mbox{sin}^{2} 2\theta$             &  1.0                   &   [ 0.93, 1.0 ]    &  [ 0.93, 1.0 ] \\ 
 $\mbox{sin}^{2} 2\bar	\theta$       &  1.0                   &   [ 0.83, 1.0 ]    &  [ 0.78, 1.0 ] \\ 
      \hline
      \hline
      \end{tabular}
      \caption{
          Best fit information for the four parameter fit to the SK-I+II+III data. 
          The 90\% C.L. column represents bounds taken from single-parameter 
          $\Delta \chi^{2}$ distributions in which the remaining three parameters have been 
          minimized over. The third column shows the 90\% C.L. allowed region when the effects of $\theta_{13}$  
          and $\delta_{\rm cp}$ are considered in the fit (see text). 
         }
    \label{tbl:fits}
   \end{center}
\end{table}

The 120 sources of systematic uncertainty considered in this analysis are separated
into two classes: those that are common throughout the SK run periods and those 
that are dependent upon a particular detector geometry. Common systematic errors 
stem from uncertainties in the neutrino interaction cross sections, nuclear effects, 
and the atmospheric neutrino flux. Independent 
systematic errors are related to detector performance and include  
uncertainties in the event reconstruction and reduction. 
A complete list of the systematic errors used here 
is presented in~\cite{Wendell:2010sd}. 

\begin{figure}[ht]
  \begin{minipage}{3.3in}
      \includegraphics[width=2.8in, keepaspectratio=true, type=pdf,ext=.pdf,read=.pdf]{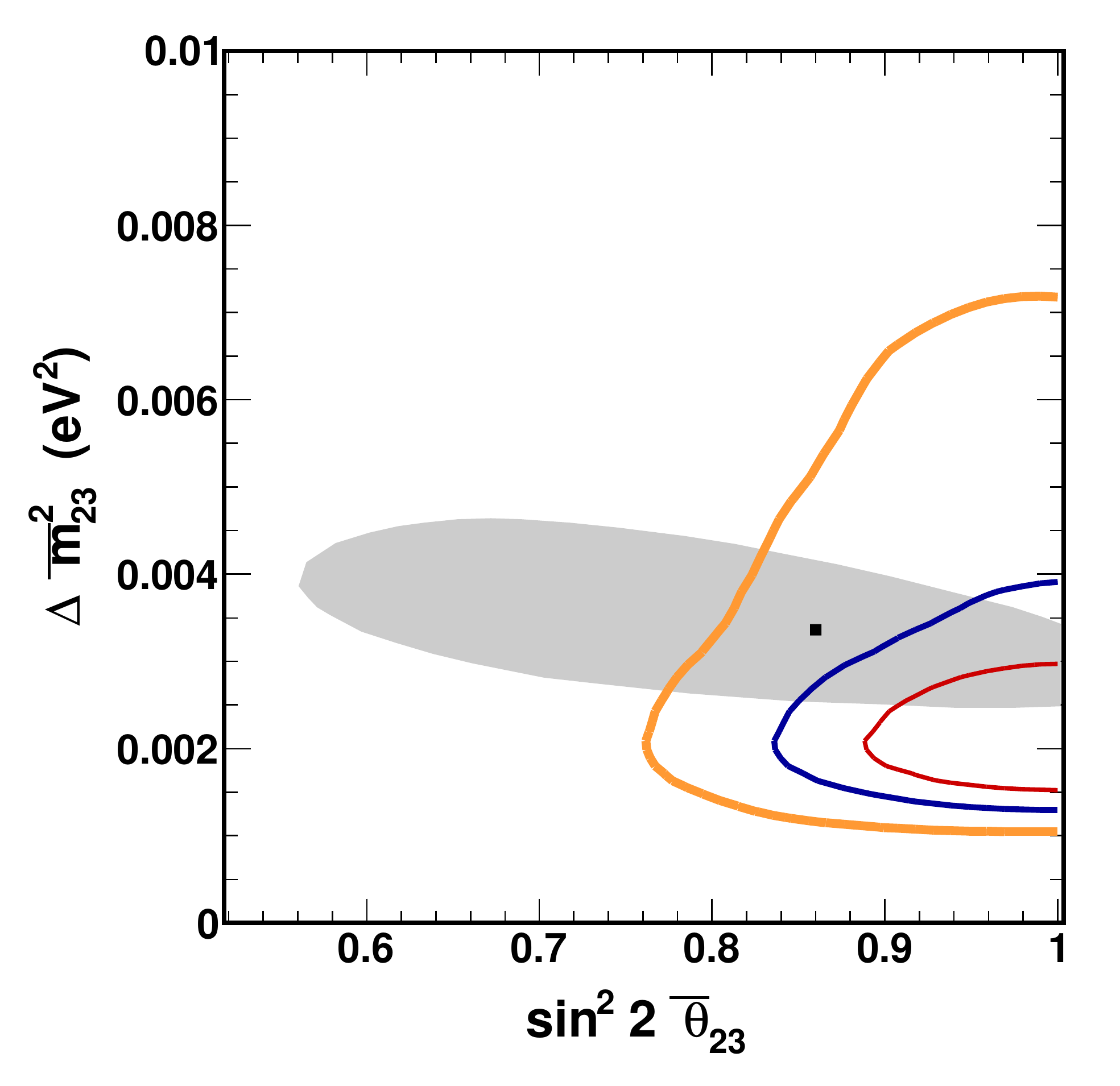}
    \caption{ Allowed regions for the antineutrino mixing parameters for the SK-I+II+III data set. 
              The 68\%, 90\%, and 99\% allowed region appear in thin, medium, and thick lines respectively. 
              The shaded region shows the 90\% C.L. allowed region 
              for antinuetrino disappearance in an antineutrino beam from MINOS~\cite{Adamson:2011fa}.
              A solid point denotes the location of the best fit from that analysis.
            }
  \label{fig:cpt.nubar.allowed}
  \end{minipage}
\end{figure}

%
Antineutrino oscillations are considered independently of neutrino oscillations 
over a four-dimensional oscillation space with two parameters for each: 
($\Delta \bar m^{2}$, $\mbox{sin}^{2} 2\bar{\theta}$) and 
($\Delta m^{2}$, $\mbox{sin}^{2} 2\theta$). 
All parameters are varied simultaneously on a grid of $50 \times 35$ points in 
the antineutrino plane and $20 \times 10 $ points in the neutrino plane. 
The neutrino(antineutrino) parameter space is taken over   
$1.0 (0.7)\times 10^{-3} \le \Delta m^{2} \le 5.0 (8.0) \times 10^{-3} \mbox{eV}^{2}$  
and
$0.85 (0.65) \le \mbox{sin}^{2} 2\theta \le 1.0 (1.0)$, 
comprising an area encompassing the current allowed values of these 
parameters~\cite{Minos08,Wendell:2010sd}. 
Minimizing the $\chi^{2}$ function in Eqn. \ref{eq:fullchi} over this parameter space,
the best fit is found 
%
%
with $\chi^{2} = 468.4$ for 416 degrees of freedom. 
Table~\ref{tbl:fits} summarizes the fit information.
Since the oscillations of atmospheric neutrinos are sensitive to the 
effects of $\theta_{13}$ and $\delta_{\rm cp}$, an additional analysis has 
been performed assuming distinct three-flavor mixing between neutrinos and 
antineutrinos. Though this model has no strong theoretical motivation, 
the potential impact of a three-flavor framework on the SK allowed regions 
is provided in the table for reference.
In both fits no difference between antineutrino and neutrino mixing is found in 
the data. 

\begin{figure}[ht]
  \begin{minipage}{3.3in}
      \includegraphics[width=2.8in, keepaspectratio=true, type=pdf,ext=.pdf,read=.pdf]{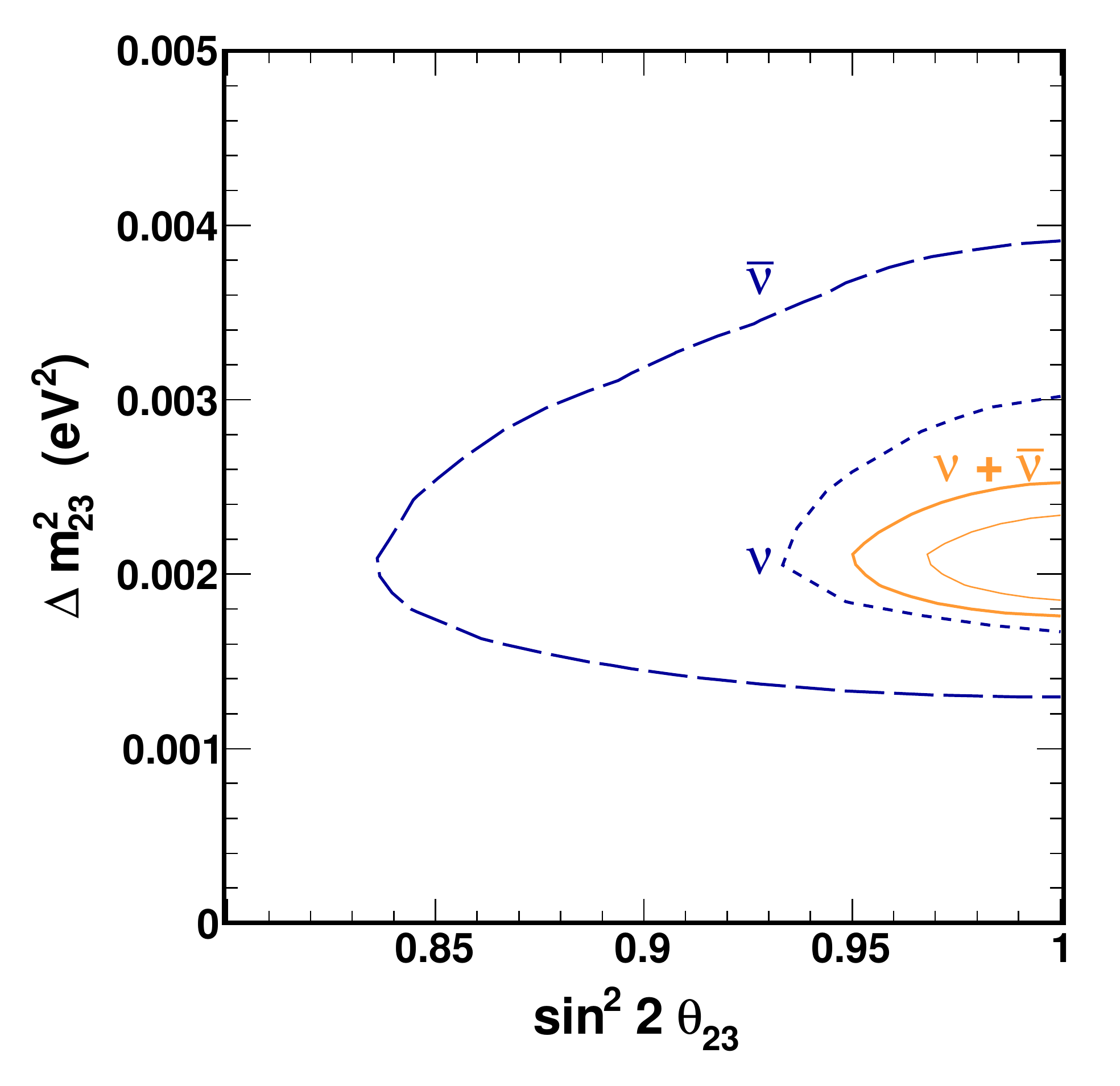}
    \caption{ 
                Allowed regions for atmospheric mixing parameters from the SK-I+II+III data set. 
                The 68\% and 90\% C.L. interval from the standard two-flavor analysis 
                with equal neutrino and antineutrino oscillations shown by 
                thin and thick solid lines, respectively. 
                The 90\% C.L. contour for neutrinos (least expansive) and antineutrinos (most expansive) 
                from the current analysis are dashed.
            }
  \label{fig:2f.cpt.overlay}
  \end{minipage}
\end{figure}

The combined data overlaid with the MC expectation at the best fit point 
are shown in Fig.~\ref{fig:cpt.zenith}. Antineutrinos have been oscillated 
independently of neutrinos in the expectation and are shown as the shaded portion of the histogram. 
Figure \ref{fig:cpt.nubar.allowed} shows the allowed regions at several C.L. in the 
antineutrino plane. The neutrino parameters have been minimized over and the 
contours have been drawn for a standard $\Delta \chi^{2}$ distribution with two degrees of freedom. 
The 90\% C.L. regions from the antineutrino and neutrino parameters overlaid 
with the allowed region from the standard two-flavor analysis, where the mixing parameters are 
required to be identical for neutrinos and antineutrinos, appear in Fig.~\ref{fig:2f.cpt.overlay}.
The best fit point of the standard analysis, 
($\Delta m^{2}$, $\mbox{sin}^{2} 2\theta$) = ( $ 2.1 \times 10^{-3} \mbox{eV}^{2}$, $1.0$~),
is consistent with the best fit from the antineutrino separated analysis.

\begin{figure}[ht]
  \begin{minipage}{3.3in}
      \includegraphics[width=2.8in, keepaspectratio=true, type=pdf,ext=.pdf,read=.pdf]{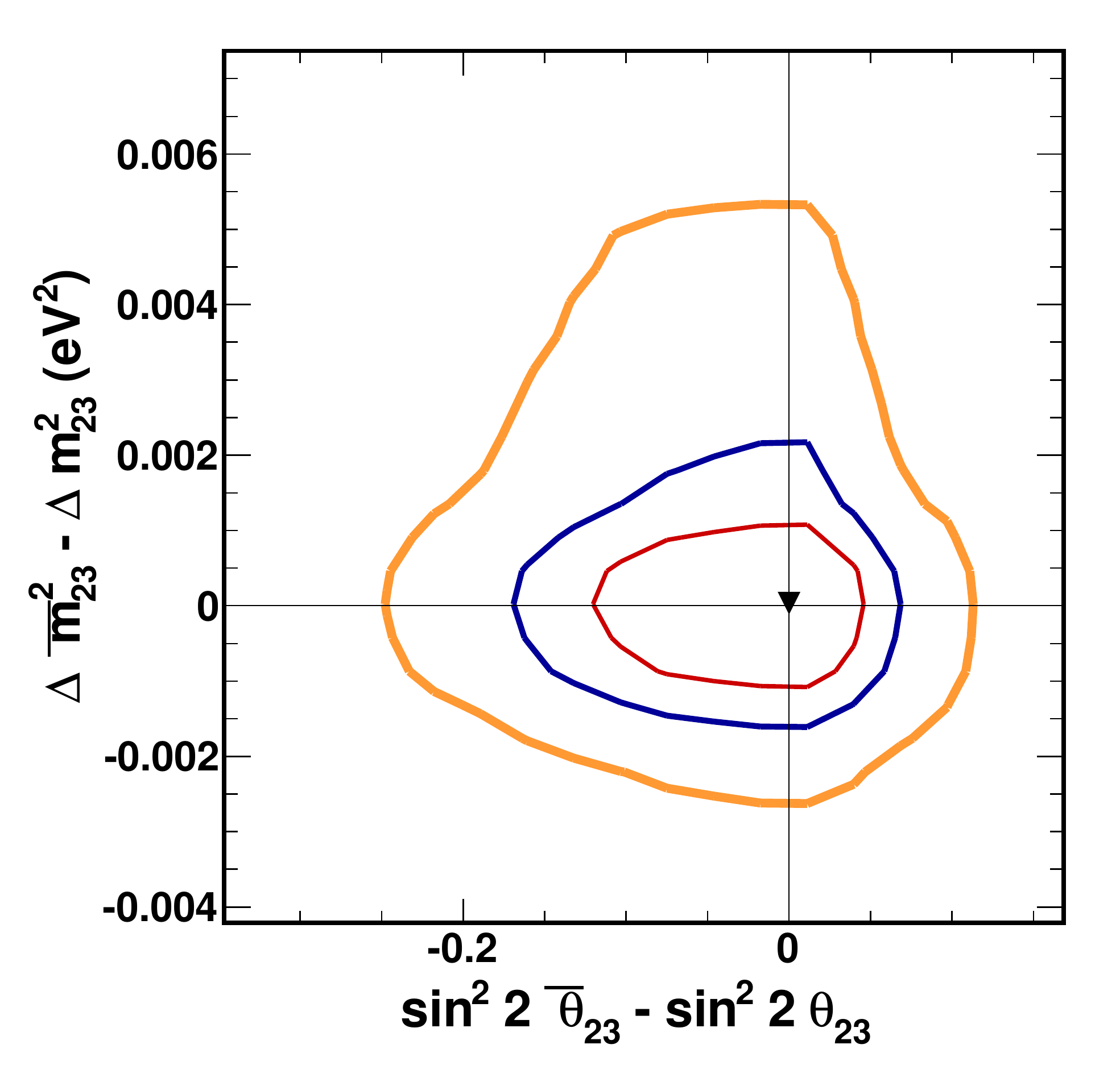}
    \caption{ Allowed differences between neutrino and antineutrino oscillation parameters for the SK-I+II+III data set. 
              The 68\%, 90\%, and 99\% allowed region appear in thin, medium, and thick lines, respectively. A black triangle 
              shows the location of the best fit point from this analysis. 
            }
  \label{fig:cpt.subtraction}
  \end{minipage}
\end{figure}

At each point in the plane of Fig.~\ref{fig:cpt.nubar.allowed} the best fit 
to the data may lie at a point in the neutrino parameters that does not 
correspond to equal neutrino and antineutrino mixing. 
To illustrate the difference between neutrino and antineutrino oscillations 
permitted by the data, Fig.~\ref{fig:cpt.subtraction} shows the allowed 
regions as a function of the difference of the antineutrino and 
neutrino mixing angles, $\mbox{sin}^{2}2\bar \theta - \mbox{sin}^{2}2 \theta$,
and mass squared splittings, $\Delta \bar m^{2} - \Delta m^{2}$. 
Contours have been drawn as in Fig.~\ref{fig:cpt.nubar.allowed}
and a black triangle near the origin represents the position of 
the best fit.

In conclusion, a search for evidence of differing neutrino and antineutrino oscillation parameters 
in the SK-I+II+III
atmospheric data sample has been carried out. The atmospheric 
mixing parameters for antineutrino oscillations are consistent with 
those for neutrino oscillation. 
The results agree with the standard SK atmospheric oscillation analysis, which have also been presented. 
The SK antineutrino oscillation best fit is consistent with the parameters found by MINOS using a predominantly muon neutrino beam.


We gratefully acknowledge the cooperation of the Kamioka Mining and
Smelting Company.  The Super-Kamiokande experiment has been built and
operated from funding by the Japanese Ministry of Education, Culture,
Sports, Science and Technology, the United States Department of Energy,
and the U.S. National Science Foundation. 
Some of us have been supported by funds from the Korean Research Foundation (BK21), and the Korea 
Research Foundation Grants (MOEHRD, Basic Research Promotion Fund), (KRF-2008-521-c00072). 
Some of us have been supported by the State Committee for Scientific Research in Poland (grant 1757/B/H03/2008/35).
Some collaborators have been supported by the National Natural Science Foundation of China under
Grants No. 10875062 and 10911140109.

\bibliography{cptv.sk1o2o3}

\end{document}